\documentstyle[11pt,newpasp,twoside,psfig]{article}
\index{summary}
\index{instructions}
\index{template}

\def\edcomment#1{\iffalse\marginpar{\raggedright\sl#1\/}\else\relax\fi}
\reversemarginpar

\begin{document}
\title{Electric Field Screening by Pairs in the Presence of Returning
Positrons}
 \author{Shinpei Shibata}
\affil{Department of Physics, Yamagata University}

\begin{abstract}
We solve the one-dimensional Poisson equation along a magnetic field line,
both analytically and numerically, for a given current density
incorporating effects of returning positrons. 
We find that
the number of returning positrons per one primary electrons
should be smaller than unity, and the returning of
positrons occurs only in a very short braking distance scale.
As a result, for realistic polar cap parameters, the accelerating 
electric field will not be screened out; thus, the model fails 
to be self-consistent. 
A previous belief that pair creation
with a pair density higher than the Goldreich-Julian density immediately
screens out the electric field is unjustified.
We suggest some possibilities to
resolve this difficulty.
\end{abstract}

\section{Introduction}

High energy pulsed emission from rotation powered pulsars
are accounted for by presuming an electric potential drop along 
a magnetic flux in the magnetosphere, as typically seen in 
`polar cap models' or `outer gap models'.
The available potential drop produced by the neutron star
is enough for the emission. In particular, for young pulsars
acceleration potential is only a small portion of the total
electromotive force. Rather important is how we understand
localization of the electric field.
In some mechanism, the actual charge density deviates from the
Goldreich-Julian (GJ) charge density to produce field-aligned 
electric field.
It is sometimes naively assumed that if pairs are created
with a density higher than the GJ density, then
produced field-aligned electric field is screened out and shut
down the accelerator localized in a small region.
In this work, we treat such a screening process by pairs and
find out conditions to obtain a finite potential drop.

Let us consider a steady model for the field-aligned accelerator which
has a finite potential drop with the electric field screened at the both ends.
For definite sign of charge, let us assume electrons are accelerated
outwards, i.e., the electric field points toward the star.
Gamma-rays  emitted by the electrons  convert into pairs
beyond a certain surface called
the pair production front (PPF);  beyond PPF pairs are assumed to 
be created continuously in space.
A schematic structure is shown in Fig.~1, and the electric potential
may be given by, in non-dimensional form, 
\begin{equation}
- {d^2 \phi \over d x^2} =
- {j \over v_1/c} + {B_z \over B} +
\bar{\rho}_- + \bar{\rho}_+
\end{equation}
where the first negative term is due to the primary electron beam,
the second term is the normalized GJ density, and the
last two terms appear due to pairs only beyond PPF.

In the `standard' polar cap model, the negative space charge (cathode)
is produced by non-relativistic electrons near the surface (when
$v_1 \ll c$). If the length of the acceleration region along field
lines is larger than the transverse scale, negative charge appearing
on the side wall also contributes.
Such an electric field can be screened by 
two possible reasons;
one is when the GJ term (the second term)
dominates the primary electronic charge (the first term), and 
the other is when pairs are created.

\begin{figure}
\begin{center}
\mbox{\psfig{file=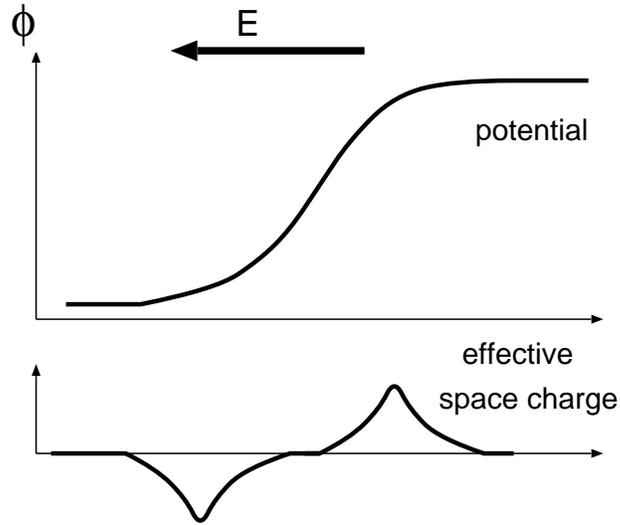}}
\end{center}
\caption{Schematic picture of the field-aligned accelerator}
\end{figure}

In the former case, it is obvious that the primary 
current should be less than a value, typically of order of GJ value,
depending how slowly $B_z$ decreases as compared with $B$.
As is well-known, field lines curving toward the rotation axis,
$B_z/B$ term increases and is favorable to the case
(Scharlemann, Arons, \& Fawley; 1978,
Arons, \& Scharlemann; 1979).
In order to shutdown the potential drop, the current density
should be adjusted to be a critical value determined by the
field geometry. However, the current density
is determined globally, especially, through interaction with the wind zone,
so the actual current density is likely to be different from the
critical current density.
The screening is nothing to do with the pair creation in this case.
If pair creation takes place, the strict condition on the 
current density is slightly relaxed; as will be shown below
some of pair positron returns back to the star and left pair
electrons behind, which provide negative space charge in the `anode region'.
Therefore if the actual primary current is {\it less} than the
critical value, i.e., if the region needs negative charge 
to adjust the GJ (positive) term, then
pair plasma helps to adjust the space charge to close the accelerator.
If the situation is opposite, namely, if current density is much higher
than the critical value, there is no way to close the accelerating electric
field, even in the away curved field lines.

I would like to argue that such super critical current density
seems realistic. The critical charge density is more or less
GJ. The GJ current is obtained by some sort of dimensional 
analysis: the current derived form the magnetic moment $\mu$,
angular velocity $\Omega$ and other physical constants
is $I = \mu^2 \Omega^4 /c$. This estimate for the circulating current
in the magnetosphere will be correct. However, the GJ current density
$B \Omega / 2 \pi c$ is just $I$ divided by the
polar cap area. It is not likely that current density is 
distributed more or less uniformly over the polar caps, and
much likely that it is
localized in annular rings or clumps, hinted by radio profiles
and also by terrestrial aurora. 
An expected current density can be by factor of 10 or 100 larger
than the GJ current density.
If one assume such a super GJ current density, it is certain 
that strong accelerating electric field appears 
as shown by Shibata (1997) 
regardless of the field geometry.

The second case that the screening is due to pairs may be sometimes
assumed.
Pair polarization is an effect that screens out the electric
field.
The pair positrons are decelerated to non-relativistic speed
soon after their birth,
while the pair electrons are accelerated,
and as a result of continuity,
a positive space charge appears to reduce the electric field.
However, some positrons reflected backward to the star by
un-screened electric field.
In this paper, we calculate the positive space charge produced by 
pair polarization and obtain the electric field strength
which can be screened.

\section{Model}

We assume one dimensional primary stream along a field line,
and pairs, which  is created beyond the PPF more or less uniformly.
Pair density is parameterized by 'multiplication factor' $M(x)$,
by which the primary electronic flux is multiplied to give 
pair flux at the point $x$. The Poisson equation for the
electric potential and equations of motion and of continuity
for electrons and positrons are solved self-consistently for 
a steady state solution.

\section{Results}

We obtain the electric field strength which can
be screened, given a pair creation rate.
It is found that returning of positrons makes the screening difficult
seriously, because the pair electrons left behind the returning positrons
produce negative space charge in the screening region where the positive
space charge is required.
As a result, the thickness of the screening is restricted to be as small as
the braking distance $\Delta s \approx mc^2 / e | E_{\parallel} |$ 
for which positrons become non-relativistic, where
$E_\parallel$ is the electric field strength just before the PPF.
We confirmed the previous result of Shibata et al. (1998) 
that the electric field
which can be screened by pair polarization is fairly small:
\begin{equation}
{E_\parallel^2 \over 8 \pi } <
mc^2 \left( \Omega B \over 2 \pi c e \right) \zeta^\prime j
\Delta M_{\rm screen}
\end{equation}
where $\Delta M_{\rm screen}$ is the pair multiplication factor
{\em within} $\Delta s$,
$\zeta^\prime$ is a numerical factor of order of unity.

If the primary current density is of order of the Goldreich-Julian (GJ) value,
the required pair multiplication factor per one primary electron is 
enormously large and cannot be realized in the conventional pair
creation models.
A previous belief that pair creation
with a pair density higher than the GJ density immediately
screens out the electric field is unjustified.

Some mechanism to salvage this difficulty should be found.
As has been mentioned in \S 1, 
the toward curvature provides an effective positive charge, so that
the required multiplication factor is considerably reduced.
In this case, over-screening by the positive GJ term 
is canceled by the pair electrons left behind the returning 
positrons (this is the opposite of pair polarization).
Harding \& Muslimov (2001) is the case, but the current is restricted to be
sub-GJ. In a more likely case with super-GJ current density,
toward curvature will not be helpful.
In any case,
physics in focused field-aligned current,
e.g., effects of inhomogeneity across the magnetic field lines,
two stream instability in a high current density with unscreened 
electric field, 
has yet been properly studied. 

Another possible way out is to include frictional forces
between various components of charged currents.
Since friction on positrons pulls them outwards
along with electrons, returning fraction of positrons will be reduced,
which will make screening easier.
Although there are some ideas for physical processes of friction
(two stream instability, production of positroniums and others),
it is not clear whether the frictional force can be strong enough
to lift positrons and screen the electric field.
Study on the frictional interaction under {\it unscreened} 
electric field is strongly demanded.

More detailed analysis and results will be published in near future
(Shibata, Miyazaki \& Takahara 2001).

\subsection*{Acknowledgments}

We thank D. Melrose and Q. Luo for useful comments and discussions.
This work is supported in part by Grant-in-Aid for Scientific
Research from the Ministry of Education, Science, and Culture of
Japan Nos. 12640229.

\end{document}